\documentclass[12pt]{article}

\usepackage{rotating} 
\usepackage{longtable}
\usepackage{array}
\usepackage{caption}
\usepackage{lscape} 

\usepackage[table,xcdraw]{xcolor}
\usepackage[table,xcdraw]{xcolor}

\usepackage[utf8]{inputenc}
\usepackage[english]{babel}

\usepackage{pdflscape}
\usepackage[a4paper,width=170mm,top=15mm,bottom =25mm]{geometry}

\usepackage[utf8]{inputenc}
\usepackage[english]{babel}
\usepackage{csquotes}
\usepackage{rotating}
\usepackage{pdflscape}
\usepackage{graphicx}
\usepackage{url}
\usepackage{array} 
\def\BibTeX{{\rm B\kern-.05em{\sc i\kern-.025em b}\kern-.08em
    T\kern-.1667em\lower.7ex\hbox{E}\kern-.125emX}}
\usepackage{caption} 
\usepackage{multirow} 
\usepackage{booktabs} 
\usepackage{array} 
\usepackage{amsmath}
\usepackage{mathtools}
\usepackage{rotating}
\usepackage[explicit]{titlesec}

\usepackage[affil-it]{authblk} 

\title{When Eye-Tracking Meets Machine Learning: A Systematic Review on Applications in Medical Image Analysis}

\author[1,2,*]{Sahar Moradizeyveh}
\author[1,3]{Mehnaz Tabassum}
\author[1,3]{Sidong Liu}
\author[1]{Robert Ahadizad Newport}
\author[1,2]{Amin Beheshti}
\author[1,*]{Antonio Di Ieva}

\affil[1]{\small Computational Neurosurgery (CNS) Lab, Macquarie Medical School, Faculty of Medicine, Health and Human Sciences, Macquarie University, Sydney, Australia}
\affil[2]{\small School of Computing, Faculty of Science and Engineering, Macquarie University, Sydney, Australia}
\affil[3]{\small Centre for Health Informatics, Australian Institute of Health Innovation, Macquarie University, Sydney, Australia}

\date{} 

\begin{document}

\maketitle

\section*{Abstract}
Eye-gaze tracking research offers significant promise in enhancing various healthcare-related tasks, above all in medical image analysis and interpretation. Eye tracking, a technology that monitors and records the movement of the eyes, provides valuable insights into human visual attention patterns. This technology can transform how healthcare professionals and medical specialists engage with and analyze diagnostic images, offering a more insightful and efficient approach to medical diagnostics. Hence, extracting meaningful features and insights from medical images by leveraging eye-gaze data improves our understanding of how radiologists and other medical experts monitor, interpret, and understand images for diagnostic purposes. Eye-tracking data, with intricate human visual attention patterns embedded, provides a bridge to integrating artificial intelligence (AI) development and human cognition. This integration allows novel methods to incorporate domain knowledge into machine learning (ML) and deep learning (DL) approaches to enhance their alignment with human-like perception and decision-making. Moreover, extensive collections of eye-tracking data have also enabled novel ML/DL methods to analyze human visual patterns, paving the way to a better understanding of human vision, attention, and cognition. This systematic review investigates eye-gaze tracking applications and methodologies for enhancing ML/DL algorithms for medical image analysis in depth.

\bigskip

\textbf{Keywords}: Eye-gaze Tracking; Medical Image Analysis; Deep Learning; Machine Learning
\section{Introduction}

Human eye-gaze tracking research extends beyond the restriction of controlled laboratory environments into day-to-day scenarios. Exploring eye tracking and gaze estimation to improve the qualitative and quantitative evaluation of eye movements during image visualization has emerged as an exciting research frontier across various disciplines, ranging from healthcare to geo-systems, aviation, computer vision, and AI. Over the years, eye-gaze tracking technology has been broadly explored in various medical applications and continues to evolve with new applications and research directions emerging over time.

The study of \textbf{\emph{eye-gaze tracking}} in medical imaging primarily involves analyzing where and how medical professionals, e.g., radiologists, look at images in different modalities, such as X-rays, ultrasound images, computed tomography (CT) scans, or magnetic resonance images (MRIs). This information can provide valuable insights into diagnostic processes and expert-related decision-making.

The early research by \emph{Kundel and Nodi}~\cite{Other_History_003} in 1978 marked a significant beginning in this field. They used eye-gaze tracking technology to study how doctors examined chest X-rays for signs of lung conditions. This pioneering work shed light on radiologists' visual and cognitive processes, providing a better understanding of how they detect abnormalities in medical images.
Following this, the work by \emph{Carmody et al.}~\cite{Other_History_004} in 1980 further expanded on these insights and discovered the influence of radiologists' eye movements, known as scan paths, on the rate of false-negative errors while examining nodules in chest X-ray images.
Since these early studies, eye-gaze tracking in medical imaging has evolved significantly. Modern technology has enabled more precise and detailed tracking of eye movements. This progress has allowed researchers to gain deeper insights into the cognitive processes of medical professionals during diagnosis, potentially leading to improved training methods and diagnostic tools.

Understanding eye movement parameters, eye tracking concepts, and their association with global-focal search models, which describe how viewers alternate between a rapid, holistic scanning of the scene and a more detailed, focused examination of specific areas of interest, is crucial to capturing the research matter. These eye movement parameters and their combined benefit in analyzing visual search patterns of radiological images and the most influential eye activities used for this purpose include fixation, saccades, scan path coverage, and the other eye movement parameters used to capture and organize complicated search patterns, known as visual search patterns.

\textbf{\emph{Fixation}} refers to preserving the visual gaze on a specific location, which facilitates the potential for the brain to receive detailed visual information from a specific point of interest. Analysis of the fixation provides insights into cognitive processes, attention, and how information is processed visually. 
Rapid eye movements, known as \textbf{\emph{saccades}}, are the eye movements that simultaneously shift the focus of eye gaze from one point to another in the same direction. Additionally, \textbf{\emph{scan paths}} represent eye movements' patterns, including a series of fixations and saccades that outline the sequential path the eyes follow as they move from one point to another while observing a scene or image. Scan paths can reveal how an individual processes visual information, highlighting the priority and sequence of the observed elements in a scene or image.
While \textbf{\emph{eye tracking}} measures a person's eye movement, activity, and position over time, providing insights from visual attention by measuring eye activities such as pupil dilation, eye blinks, fixations, and saccadic movements; \textbf{\emph{gaze tracking}} represents a detailed analysis of eye tracking data, aiming to estimate the direction of a person's gaze and interpret their interactive engagement with a given scene. 
In healthcare applications, although the differences in cognitive processes and perceptual patterns during image analysis can lead to diverse interpretations of the same image, the dependence on diagnostic determinations in visual search patterns offers a valuable chance to incorporate this detailed supplementary information into computer-assisted diagnosis systems. Hence, tracking radiologists' and other medical experts' eye movement strategies while reading the medical images shows significance in lesion detection and disease diagnosis. The research in this field has garnered considerable interest and attention in recent years, reflecting its growing importance and potential impact on various fields.

Accurate visual data interpretation is significant for diagnostic and decision-making in medical fields such as radiology, pathology, dermatology, ophthalmology, and neurology, among others, which is directed to growing attraction with the potential of eye gaze patterns to delineate the cognitive abilities of specialists or to forecast and identify disorders ~\cite{last}. In this review, we aim to investigate the use of human visual attention in interpreting medical images to enhance the performance of Machine learning and deep learning methodologies used in medical image analysis. Based on the findings in current research, we classified the eye-gaze tracking applications with ML/DL approaches into five principal clusters for identifying the search patterns, facilitating decision-making, providing educational and training resources, accelerating disease detection to diagnosis and classification, and assessing expertise fatigue and skill levels. Moreover, the objective is extended to analyze their applications in different tasks, including medical image annotation, classification, pathology detection, and segmentation. Figure 1 shows the general approaches and phases that leverage eye-gaze tracking data to enhance the performance of the ML/DL methods.
\begin{figure*}[!ht]  
    \centering
    \includegraphics [scale=0.29]{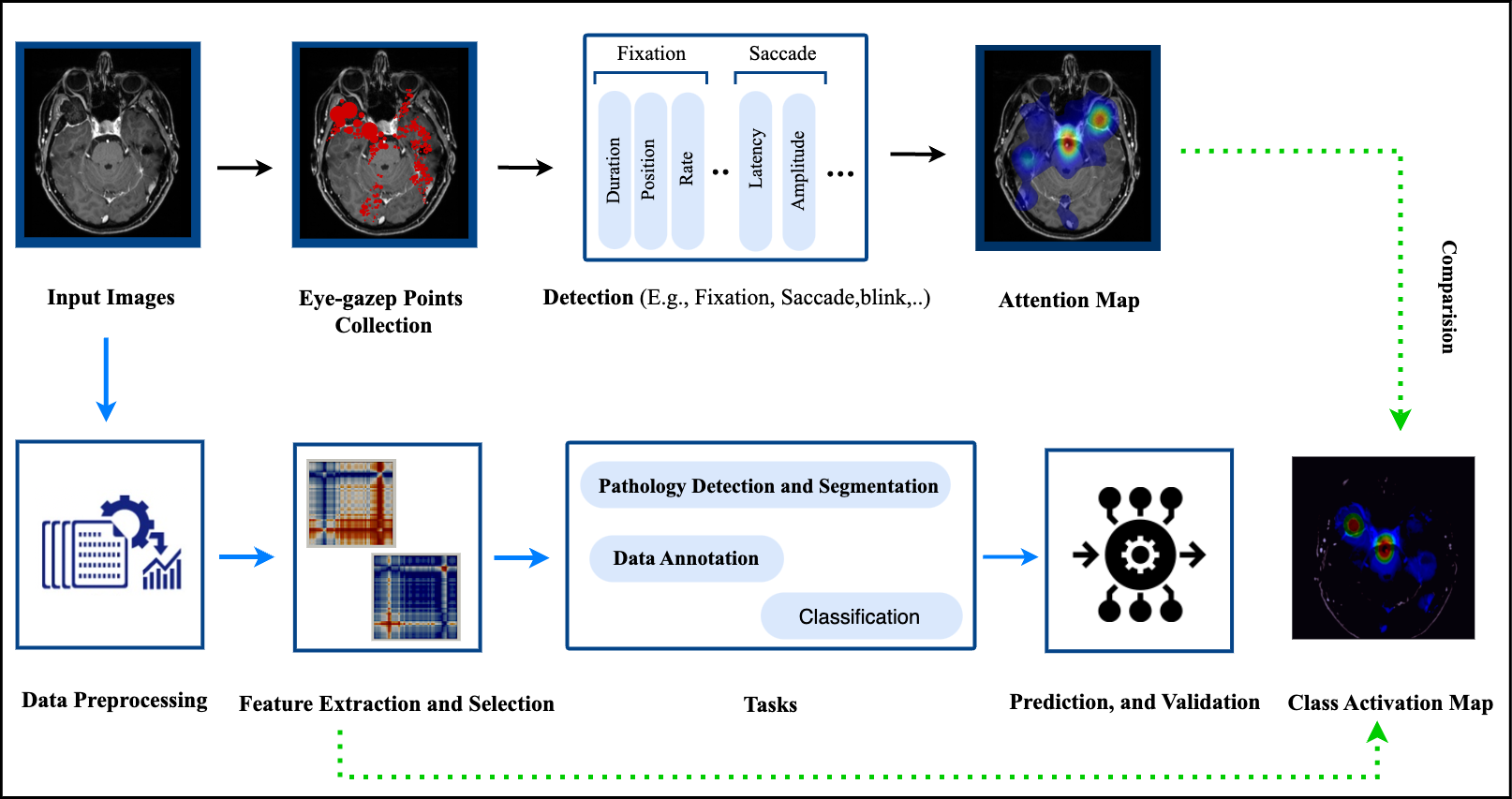}
    \caption{\footnotesize An integrated ML/DL pipeline utilizing eye-gaze tracking data to improve diagnostic accuracy in medical imaging. For example, the input image in this example shows an axial slice of a brain MRI with a skull Meningioma seen by a cohort of neuroradiologists and neurosurgeons (experts’ scan paths and their analysis). In additional steps, features are extracted from the MRI images and scan paths for various tasks. }
    \label{fig:method}
\end{figure*}
In this review, we aim to investigate the use of human visual attention in interpreting medical images to enhance the performance and productivity of deep learning algorithms. Therefore, based on the findings in current research, we classified the eye-gaze tracking applications with ML/DL approaches into five principal clusters for identifying the search patterns, facilitating decision-making, providing educational and training resources, accelerating disease detection to diagnosis and classification, and assessing expertise fatigue and skill levels. Moreover, the objective is extended to analyze their applications in different tasks, including medical image annotation, segmentation, object detection, and classification. 

\section{Materials and Methods}

\subsection{ Literature Selection and Search Strategy}

A comprehensive literature review was performed using the Preferred Reporting Items for Systematic Reviews and Meta-Analyses (PRISMA) guidelines framework. Ethics approval was not necessary for this review. Two assessors identified all studies that could contribute to addressing at least one aspect of our research inquiries. The SLR includes all related articles in English between January 2018 and December 2023 in Google Scholar and PubMed databases. We did our best to have an unbiased search strategy that contained all potential search strings associated with the topic. The following search strings, using \textbf{AND} and \textbf{OR} operators, were functions on both databases to indicate the generic search queries.\newline

\textbf{((EYE-GAZE) AND (TRACKING OR ESTIMATE OR PREDICT) AND ("MEDICAL IMAGE") AND (RADIOLOGIST OR RADIOLOGY) AND ("DEEP LEARNING" OR "MACHINE LEARNING" OR "NEURAL NETWORKS")} \newline

The focus of this research is on the assessment of applying ML and DL methodologies to analyze eye-tracking data derived from medical practitioners. Additionally, it explores how eye-tracking data can be utilized to improve ML/DL algorithms for medical image analysis. An initial comprehensive search across both academic databases yielded 299 articles. After removing duplicates, 267 unique articles remained. These articles were studied further by reviewing their abstract and excluding 189 from the study, relying on the following: Review paper, book chapter, non-gaze studies, non-gaze tracking on medical images, not ML/DL study, and unrelated research. Seventy-eight papers were analyzed for more details. Ultimately, 31 articles related to eye-gaze tracking in medical image analysis were explored in this paper, which is demonstrated in Figure 2.

\begin{figure*}[!ht]  
    \centering
    \includegraphics [scale=0.5]{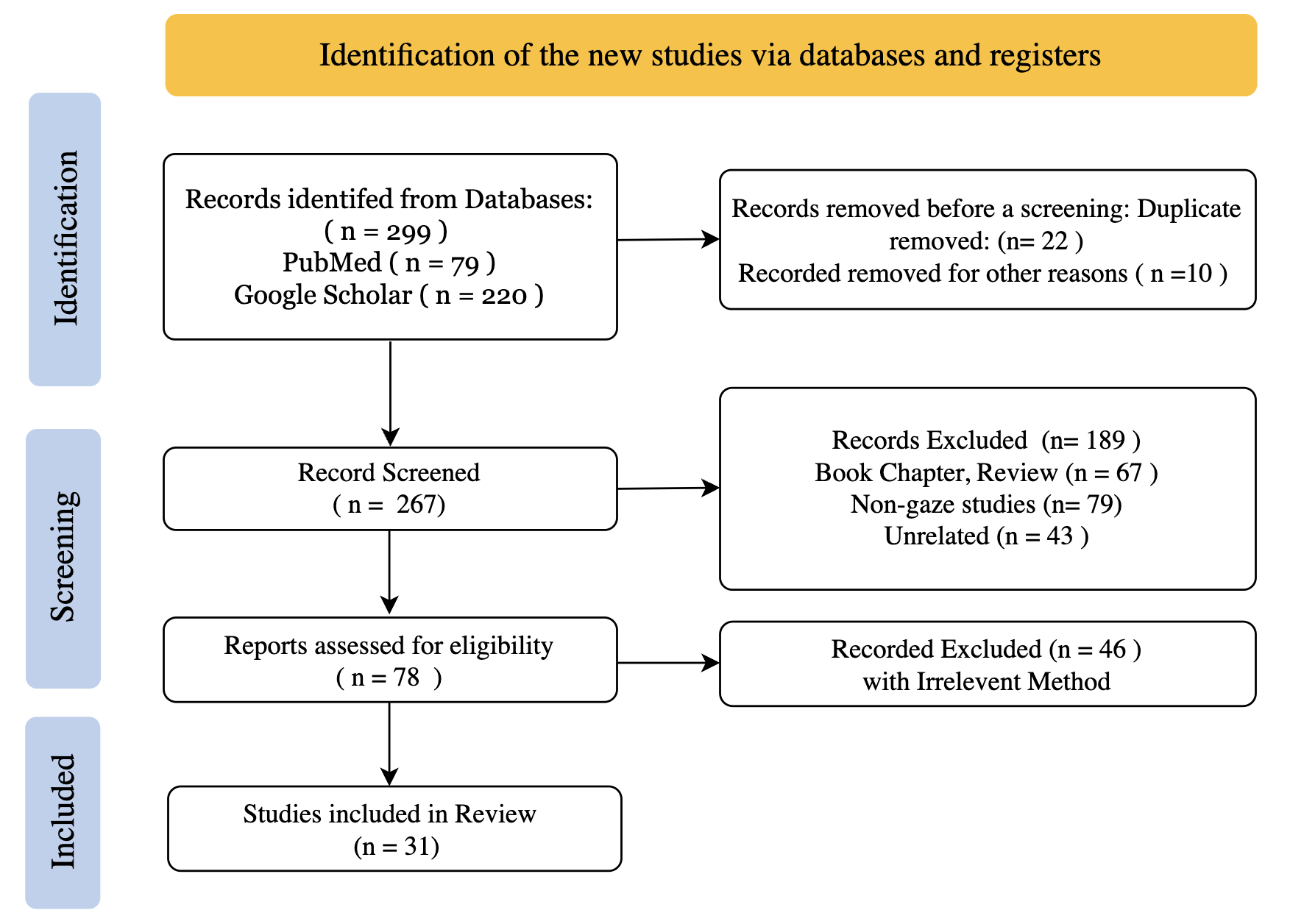}
    \caption{\small Overview of PRISMA chart in Eye-Gaze Tracking using ML/ DL in Radiological Medical Image Analysis.}
    \label{fig:Prisma}
\end{figure*}

\subsection{ Quality Assessment}
The Quality Assessment of Diagnostic Accuracy Studies-2 (QUADAS-2) tool, a robust and dependable instrument, is used in this review paper. As illustrated in Figure 3, the tool comprehensively assesses each study. In the evaluation of bias risk related to data selection, a considerable 83\% of the studies (26 studies) demonstrated a low risk, 6.8\% (2 studies) exhibited a high risk, and the remaining 9.6\% (3 studies) had an unclear risk. The risk of bias associated with the index test was categorized as unclear in only 9.6\% of the studies (3 out of 31), indicating a substantial majority of 90.4\% (28 out of 31) with a low risk. Similarly, for the reference-standard test, a low risk of bias was identified in 74.2\% of the studies.
In comparison, the risk level remained uncertain at 25.8\%. In the domain of process and timing, the evaluation indicated that 90.4\% of the studies (28 studies) had a low risk of bias, and the risk remained unclear in 9.6\% (3 studies). The charts show that the study has low concerns regarding its applicability to the research objective. The diagnostic tests reviewed in the study are deemed appropriate, and how the tests were conducted and interpreted is consistent with the proposed use case.

\begin{figure*}[!ht]  
    \centering
    \includegraphics [scale=0.4]{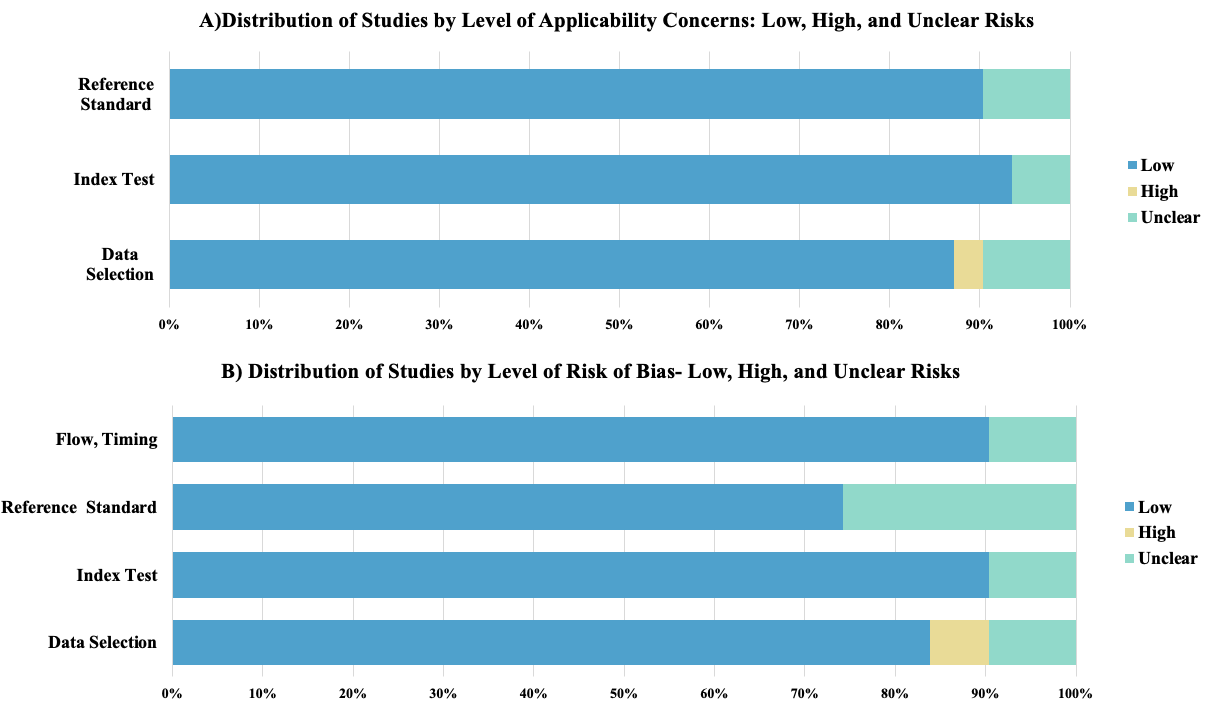}
    \caption{\small Summary of QUADAS-2 assessments of included studies.}
    \label{fig:QA}
\end{figure*}
\subsection{Characteristics of the studies}

Employing ML/DL algorithms to analyze the eye-gaze patterns of medical experts is a significant opportunity to enhance patient assessments. This approach promises more accurate, comprehensive evaluations, ultimately contributing to improved patient outcomes. We reviewed the selected articles and compiled essential information and details of the study into a structured datasheet. We extract features from each paper, including the modality, anatomical regions of interest, input type, dataset, tasks, the method employed, and the article's aims. Most articles noted the eye tracker device and the number of experts that assisted them with the research; however, some research did not mention the details. Meanwhile, if available, we collected other details, such as model design, visual attention-learning approaches, performance evaluation, and validation methods.

Figure 4. (A) presents the distribution of the eye-tracking tasks. It includes annotation at 16\%, detection at 19\%, segmentation at 16\%, and a notably more significant segment for classification, which occupies 49\%. This distribution underscores a marked trend toward increasing eye gaze data utilization, particularly in lesion and disease diagnosis and classification, indicating a growing reliance on this technology in medical diagnostics. Figure 4. (B) illustrates the distribution of various techniques referenced in the articles under study. This chart represents the quantitative aspects of different methods and indicates a considerable focus on convolutional neural networks (CNNs). Besides, the statistics point to the recent shift towards transformer-based approaches, and their strategy is to either augment or replace parts of CNN architectures and use them in conjunction with CNNs or apply them as a standalone framework. Finally, Figure 4. (C) shows the remarkable growth surge and the practical importance of the field's growth. This bar chart emphasizes the expanding relevance of the subject matter in recent times. 

\begin{figure}[ht]
    \centering
    \includegraphics[angle=0, scale=0.48]{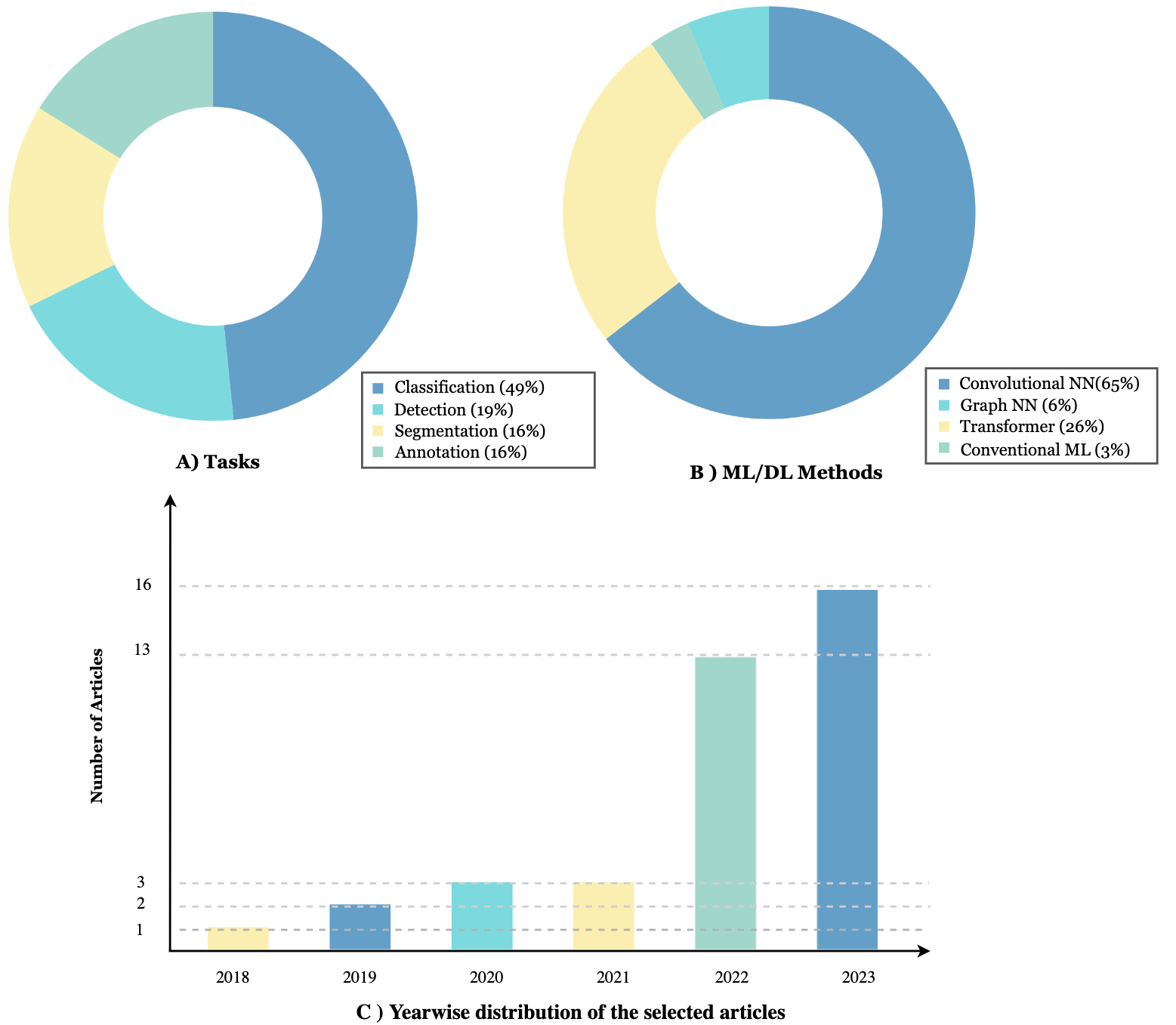}
    \caption{\small Illustrate the detailed analysis of the various eye-tracking application tasks (A) and distribution of various techniques referenced in the articles under study(B) from 2018 to 2023(C).}
    \label{fig:method-task}
\end{figure}

This review investigates the existing literature in eye-gaze tracking research with ML/DL approaches for medical image analysis. First, we studied the eye-gaze tracking data in medical imaging, including the data collection, available datasets, and eye-tracking tools and devices. Then, based on the current research, we categorized the eye-gaze tracking applications in healthcare into image classification, model interpretability enhancement, data annotation, pathology detection, and segmentation and assessed expertise workload and skills. Then, we discussed visual attention learning in interpreting and understanding the images. In the next phase, we explore the various ML/DL techniques used in this field, and we generally classify them into conventional machine learning algorithms, Convolutional Neural Networks (CNN), transformer-based methods, and approaches that leverage Graph Neural Networks (GNN). Finally, we discussed the different visual attention learning approaches used in literature for analyzing eye gaze tracking data. This comprehensive review analyses 31 articles, investigating their fundamental aspects customized to various elements. Table 1 shows the key characteristics of these articles, including imaging modalities, targeted organs, input types, applications, methodologies, eye-tracking devices utilized, the involvement of expert collaborators, primary research objectives, and publication years.

\begin{landscape}

\centering
\scriptsize
\renewcommand{\arraystretch}{1.5} 
\begin{longtable}{
  p{2.6cm}
  >{\centering\arraybackslash}p{1.5cm} 
   >{\centering\arraybackslash}p{1.5cm}
   >{\centering\arraybackslash}p{1.5cm}
   >{\centering\arraybackslash}p{1.5cm}
    >{\centering\arraybackslash}p{4cm}
  p{8cm}
  >{\centering\arraybackslash}p{1.5cm} 
  p{1.5cm}
}
\caption{\small AN OVERVIEW OF THE MAJOR EYE-GAZE TRACKING APPLICATIONS IN MEDICAL IMAGING FROM 2018 TO 2023.}\\

\toprule

\textbf{References} &
  \textbf{Modality} &
  \textbf{Organ} &
  \textbf{Input Type} &
  \textbf{ER-Device} &
  \textbf{Number of Experts} &
  \multicolumn{1}{c}{\textbf{Goal}} &
  \hfill \textbf{Year} \\ \midrule
\endfirsthead

\multicolumn{9}{c}%
{{\bfseries \tablename\ \thetable{} -- continued from previous page}} \\
\toprule
\textbf{References} & \textbf{Modality} & \textbf{Organ} & \textbf{Input Type}  & \textbf{ER-Device} & \textbf{Number of Experts} & \multicolumn{1}{c}{\textbf{Goal}} & \textbf{Year} \\ 
\midrule
\endhead

\midrule
 \\
\endfoot
\bottomrule
\endlastfoot

\multicolumn{9}{c}{\cellcolor[HTML]{FFC000}\parbox[c][3em][c]{\linewidth}{\centering\textbf{\footnotesize Data Annotation}}} \\


Wang et al. ~\cite{S_010} &
  Xray &
  Knee &
  2D &
  Tobii &
1 Radiologist &
  Integrate radiologist eye data into CAD systems to enhance their performance. &
  \hfill 2022 \\

Ji et al. ~\cite{S_019} &
  Mammogram &
  Breast &
  2D &
  Tobii &
  1 Radiologist &
  Mammo-Net integrates radiologists’ gaze data and interactive information between CC-view and MLO-view to enhance diagnostic performance. &
  \hfill 2023 \\

Teng et al. ~\cite{S_022} &
  Ultrasound &
  Fetal &
  2D &
  Tobii &
  10 Expert Operators &
  Standardize the reference for comparing eye-tracking data for skill characterization and utilize an affine transformer network to normalize this data. &
  \hfill 2022 \\

Stember et al. ~\cite{PMC_003} &
  MRI (T1) &
  Brain &
  2D &
  Fovio&
  1 Radiologist &
  Proof that the segmentation masks generated using eye-tracking technology are similar to those obtained by manual hand annotation. &
  \hfill 2019 \\

Hsieh et al. ~\cite{S_015} &
  X-ray &
  Lung &
  2D &
  Eyelink, Gazepoint GP3 &
6 Radiologists&
  Explore the integration of various data modalities into deep learning models,and reflecting how expert scan paths operate and make diagnoses. &
  \hfill 2023 \\
\multicolumn{9}{c}{\cellcolor[HTML]{FFC000}\parbox[c][3em][c]{\linewidth}{\centering\textbf{\footnotesize Object Detection}}} \\
Mariam et al. ~\cite{S_020} &
  Histopathology &
  Oral region &
  2D &
  Gazepoint GP3 &
  2 Pathologists &
  Explores the viability and timing comparisons of eye gaze labeling compared to conventional manual labeling for training object detectors. &
  \hfill 2022 \\
  
Watanabe et al.~\cite{PMC_004} &
  X-ray &
  Lung &
  2D &
  Gazepoint GP3 &
  1 Radiologist &
  Enhance the accuracy and explainability of deep learning models in radiology by focusing on integrating heatmap generators and eye-gaze data into the training process. &
  \hfill 2022 \\

Lanfredi et al. ~\cite{PMC_006} &
  X-ray &
  Lung &
  2D &
  Eyelink &
  5 Radiologists &
  Enhance the interpretability of CNNs used for chest X-ray image analysis without affecting their image-level classification performance. &
  \hfill 2023 \\

Bhattacharya et al. ~\cite{S_013} &
  X-ray &
  Chest &
  2D &
  Eyelink &
  5 Radiologists &
  Develop a method combining radiologists' eye-gaze patterns with radiomics features to improve the detection and diagnosis of diseases from chest radiographs. &
  \hfill 2022 \\

Lu et al.~\cite{S_016} &
  X-ray &
  Lung &
  2D &
  Eyelink  &
  5 Radiologists &
  Proposed Multimodal DL architecture that combines fixation maps with chest X-ray images and performs abnormality detection. &
  \hfill 2023 \\

Stember et al.~\cite{PMC_005} &
  MRI  (T1) &
  Brain &
  2D &
  Fovio &
  1 Radiologist &
  Combining eye tracking and speech recognition to automatically extract lesion location labels for deep learning, achieving a high accuracy level of 92\%.&
  \hfill 2020 \\
  
\multicolumn{9}{c}{\cellcolor[HTML]{FFC000}\parbox[c][3em][c]{\linewidth}{\centering\textbf{\footnotesize Segmentation}}} \\
Pershin et al. ~\cite{S_001} &
  X-ray &
  Lung &
  2D &
  Tobii &
  4 Radiologists &
  To determine if AI can assist in detecting changes in radiologists' gaze patterns that correlate with fatigue and evaluate how it changes with increasing workloads. &
  \hfill 2022 \\
Pham et al. ~\cite{S_003} &
  X-ray &
  Lung &
  2D &
  Eyelink &
  5 radiologists &
  Develop a unified, controllable, and interpretable pipeline for decoding the intense focus of radiologists in CXR diagnosis. &
  \hfill 2023 \\

Wang et al. ~\cite{S_009} &
  CT Scan &
  Multi-organ &
  2D/3D &
  \_\_ &
  -- &
  Develop a network to reduce the dependence on manual annotations. &
  \hfill 2023 \\

Pershin et al. ~\cite{PMC_001} &
  X-ray &
  Lung &
  2D &
  Tobii &
  4 Radiologists &
  Quantitatively assess changes in radiologists' image reading patterns about radiological workload by randomized experiment. &
  \hfill 2023 \\

Wang et al. ~\cite{S_017} &
  X-ray &
  knee &
  2D &
  Tobii &
  \_\_ &
  Improve the quality of contrastive views in medical images and model their visual attention when diagnosing X-ray images and predicting visual attention for new images. &
  \hfill 2023 \\
\multicolumn{9}{c}{\cellcolor[HTML]{FFC000}\parbox[c][3em][c]{\linewidth}{\centering\textbf{\footnotesize Classification}}} \\
Martinez et al. ~\cite{S_002} &
  X-ray &
  Lung &
  2D &
  Tobii&
  2 Radiologist &
  Address the issue of perception-related errors, which constitute a significant portion of diagnostic mistakes in radiology. &
  \hfill 2023 \\
  
Castner et al. ~\cite{S_004} &
  OPT &
  Oral region &
  2D &
  SMI&
  57 Dental students+30dental experts &
  Use eye movement data to effectively differentiate levels of expertise, particularly in medical settings like dentistry. &
  \hfill 2023 \\

Rong et al. ~\cite{S_005} &
  X-ray &
  Lung &
  2D &
  Tobii &
  CUB-GHA: 5 specialists,and CXR-Eye: 1 radiologist &
  validate the hypothesis that human attention can significantly benefit classification models. &
  \hfill 2021 \\

Akerman et al. ~\cite{S_006} &
  OCT &
  Eye &
  2D &
  Pupil Labs Core eye tracker &
  13 ophthalmology &
  To evaluate how eye gaze metrics and gaze patterns evolve with increasing levels of ophthalmic education and enhance decision-making process in ophthalmology. &
  \hfill 2023 \\

Kim et al. ~\cite{S_007} &
  X-ray &
  Lung &
  2D &
  Gazepoint GP3  &
  a Radiologist &
  Analysis of the effectiveness and reliability of saliency methods in providing meaningful insights into the model's decision-making process. &
  \hfill 2021 \\

Ma et al. ~\cite{S_008} &
  X-ray &
  Lung &
  2D &
  Tobii &
  3 Radiologist &
  Effectively rectify harmful shortcut learning and improve the interpretability of the model by introducing the model named  EG-ViT. &
  \hfill 2023 \\

Zhu et al.~\cite{S_018} &
  X-ray &
  Lung &
  2D &
  Gazepoint GP3 &
  1 Radiologist &
  Demonstrate how human visual attention can enhance machine learning models, particularly in medical tasks. &
  \hfill 2022 \\

Wang et al. ~\cite{S_012} &
  X-ray &
  Lung &
  2D &
  Gazepoint GP3 &
  1 Radiologist &
  Develop an efficient, real-time algorithm for disease classification in chest X-rays that leverages raw eye-gaze data without the need for conversion into VAMs. &
  \hfill 2023 \\

Agnihotri et al. ~\cite{S_024}&
  X-ray &
  Lung &
  2D &
  Gazepoint GP3 &
  1 Radiologist &
  Perform an exhaustive analysis using the Eye-Gaze dataset to evaluate the impact of different input features on the performance and explainability of DL classification models in radiology. &
  \hfill 2022 \\

Cao et al. ~\cite{S_025} &
Ultrasound &
 musculoskeletal system &
2D &
Tobii &
\_\_ &
Determine where, what, and how to adjust the model to emphasize necessary regions and classify instances based on reasoning by developing a framework and a training mechanism for balanced accuracy and attention to reasonability. &
 \hfill 2023 \\

Khosravan et al.~\cite{PMC_002}&
  CT and MRI &
  Lung and Prostate &
  3D &
  Fovio&
  3 Radiologists &
  By integrating eye-tracking data with a CAD system, the researchers aim to enhance the diagnostic process. They hypothesize that combining the strengths of radiologists and CAD systems will improve screening and diagnosis performance. &
  \hfill 2018 \\

Antunes et al. ~\cite{S_014} &
  PET scan &
  Brain &
  2D &
  Fovio&
  1 Medical doctor(nuclear medicine) &
  Determine where, what, and how to I33adjust the model to emphasize necessary regions and classify instances based on reasoning by developing a framework and a training mechanism for balanced accuracy and attention to reasonability. &
  \hfill 2022 \\

Bhattacharya et al. ~\cite{S_021} &
  X-ray &
  Lung &
  2D &
  Gazepoint GP3 &
  5 Radiologists &
  RadioTransformer is a framework that integrates radiologists' gaze patterns into a student-teacher transformer model for disease diagnosis in chest radiographs. &
  \hfill 2022 \\

Teng et al. ~\cite{S_023} &
  Ultrasound &
  Breast &
  2D &
  Tobii&
  14 Fully Qualified sonographers+ trainees, &
  Classify human skills in fetal ultrasound scanning using eye-tracking and pupillary data from sonographers, challenging traditional skill groupings based on professional experience. &
  \hfill 2023 \\

Zhu et al. ~\cite{S_011} &
  X-ray &
  Lung &
  2D &
  Gazepoint GP3  &
  1 Radiologist&
  Attempt to improve visual interpretability, create a lightweight and generic application, and 
enhance classification performance. &
  \hfill 2022 \\ 

  \bottomrule
\end{longtable}
\end{landscape}
\newpage
\section{Eye-gaze Tracking Datasets and Devices in Medical Imaging} 

Current eye gaze diagnostic frameworks in healthcare primarily focus on the semantic aspects of images, emphasizing that integrating this valuable domain knowledge data and eye-tracking technology provides a comprehensive data stream and supports many medical hypotheses for overcoming the complexities associated with perceptive cognitive activities.
Eye-gaze data emerges in temporal and static structures; as \emph{Agnihotri et al.} discussed in their article ~\cite{S_024}, the temporal data consists of a series of statistical heatmaps that show where radiologists focus their attention on an image during the reading process. In contrast, static data takes this information (which is dynamic and changes over time) and shrinks it into a more fixed, non-changing format. The pixel intensity within these static heatmaps reflects the duration or extent of the radiologist's gaze on each specific pixel. In other words, it quantifies how long a radiologist's attention remains on various parts of the image, helping identify regions of significant interest or concern. Different modalities, including X-ray, positron emission tomography (PET), MRI, mammogram, optical coherence tomography (OCT), ultrasound, CT scan, and histopathology images, are used to collect data for analyzing the eye gaze patterns of medical specialists.
Figure 5 summarizes human organs and the image modalities featured in eye-gaze datasets. The result highlights that chest X-rays (CXR), mainly used to assess lung conditions, are the most represented in these datasets. 

\begin{figure}[!ht]
    \centering
    \includegraphics[angle=0, scale=0.38]{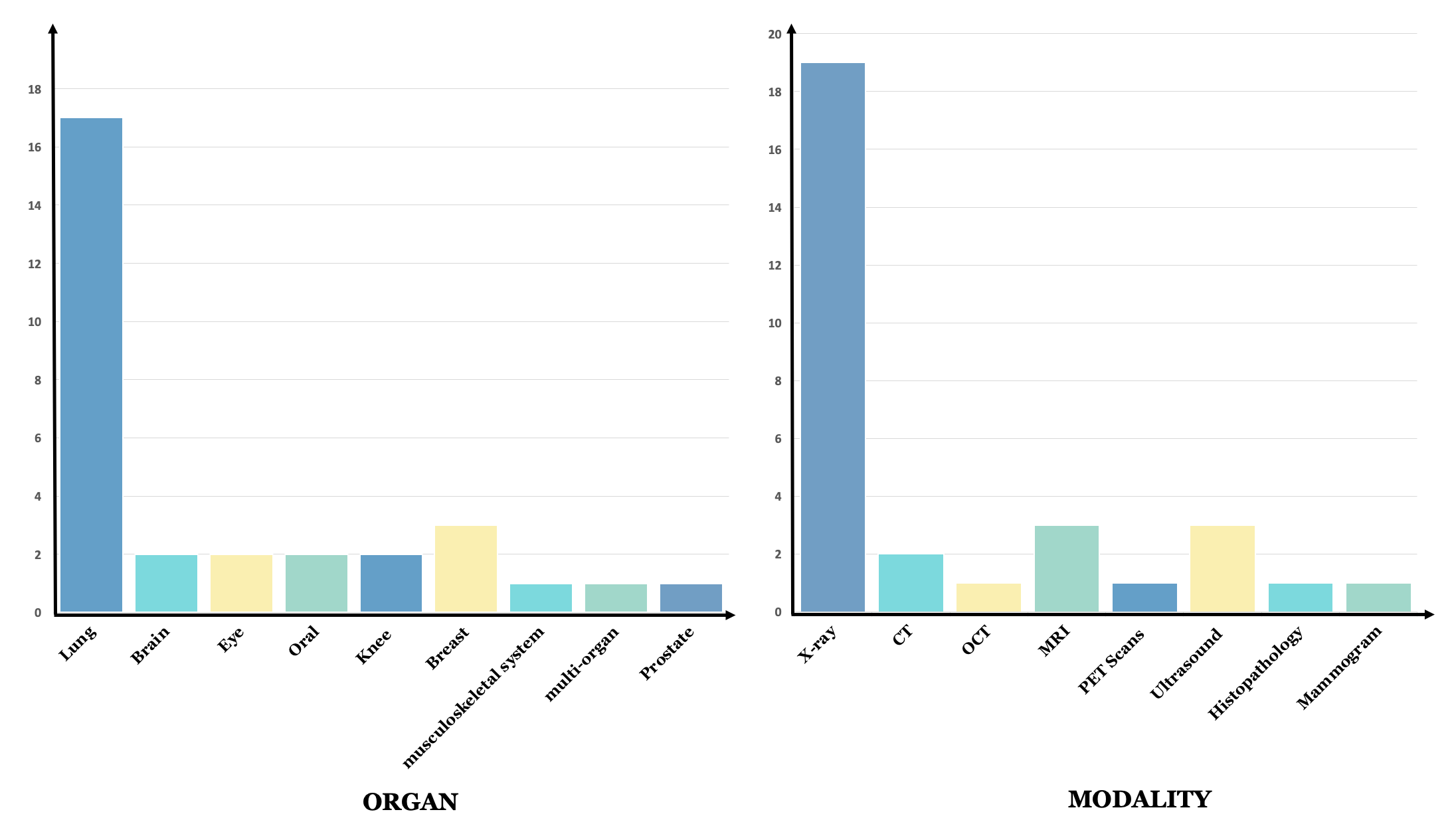}
    \caption{\small Human organs and the image modalities featured in eye-gaze datasets from 2018-2023}
    \label{fig: organ, modality}
\end{figure}

Generally, collecting eye movement data is a challenging task. Practical datasets for eye-gaze tracking research in the context of medical images are restricted due to various factors, including clinical facility, the eye tracker equipment, privacy, and ethical concerns, reliability of dataset details, and a lack of radiologists and their different scaling approaches. Therefore, only a few public datasets are available for eye-gaze tracking in medical imaging. While limited in sample size, these datasets provide valuable resources for research and development in this area.
\emph{Karargyris et al.(2021)}~\cite{Dataset_PMC_002} developed an eye-gaze tracking dataset, known as the CXR-EYE dataset founded on the MIMIC CXR dataset. A single radiologist's gaze data was collected using Gazepoint GP3 Eye Tracker. This pioneering work has since been a foundational resource for numerous research studies, significantly advancing the understanding and application of eye tracking in radiological image analysis.
L\emph{Lanfredi et al.(2022)}~\cite{Dataset_PMC_001} proposed another dataset named REFLACX. The dataset originated from the MIMIC-CXR dataset, in which five radiologists actively read, annotate, and label the chest X-rays in a structured three-phase process. They employ the Eye Link 1000 Plus eye-tracking device to capture eye-tracking data and simultaneously record their audio reports, creating a rich and detailed dataset.
\emph{Hsieh et al. (2023)}~\cite{Dataset_S_001}create a dataset named MIMIC-EYE that includes 3,689 tuples of CXR images. By integrating the chest X-ray images, radiologists' eye movement data, and patient clinical information, this dataset is designed to facilitate the development of deep learning models that can leverage multiple data types for improved diagnostic accuracy in medical imaging.
\emph{Ji et al.(2022)}~\cite{S_005} attempted to capture radiologists' scan paths and gaze patterns and construct a model for breast cancer diagnosis by using CBIS-DDSM and INbreast dataset along with Tobii Pro Nano eye tracker. Furthermore, there are specialized eye gaze datasets designed for diverse applications, which are distinct from medical images; for instance, the CUB-GHA dataset (2022) [26] focuses on fine-grained classification tasks using images of birds, coupled with eye gaze data from multiple observers to analyze human attention patterns.

Eye tracker devices observe, capture, and measure eye movements, gaze, and pupil dilation across diverse fields. These devices range from simple and affordable systems to advanced and high-precision ones, which can be chosen based on the types of stimuli, participants, testing environment, and required metrics. 
The current eye-tracking techniques include four distinct groups. \emph{Head stabilization}, which employs an approach to constrain the participants' head movements to enhance the analysis accuracy, can be used with secondary technology simultaneously.  

The remote automatically adjusts the camera field by measuring the pupil center and cornea reflection without reaching the participant's head. Head-mounted or mobile eye tracker, including wearable components such as glasses or headbands, by capturing the entire field of view of the participants, facilitates the creation of ideal experiments for real-world scenarios, and includes applications such as simulators for motor-control systems, gaming, and training. Moreover, embedded or integrated eye trackers, incorporated in virtual and augmented reality devices, with a controllable stimulus and immersive screen, function like remote eye trackers. Eye trackers are sophisticated devices, and there are many other features to characterize them, such as accuracy, precision, sampling frequency, latency, portability and wearability, calibration process, and cost. It is essential to consider the balance between these factors against the specific needs of the research or application scenario. The most frequently mentioned eye tracking devices in medical image analysis to examine the gaze pattern details and medical diagnosis are Tobii (34\%), Gazepoint GP3 (28\%), Fovio (13\%), Eye link (16\%) and SMI, Pupil lab core eye tracker (6\%). One research paper did not mention the eye tracker device used in their experiment. We also displayed the eye tracker devices used in each article in the characterization table.
\section{Eye-gaze Tracking Applications in Medical Imaging}
Eye-gaze tracking impacts medical image interpretation, analysis, and diagnosis. Utilising eye-gaze data across diverse applications leads models to focus on critical elements, objects, and specific details vital for different tasks. Accurate visual data interpretation is significant for diagnostic and decision-making in medical fields such as radiology, pathology, neurology, and others. 
In this section, we aim to investigate the use of human visual attention in interpreting medical images to enhance the performance and productivity of ML/DL algorithms. Based on the findings in current research, we classified the eye-gaze tracking applications with ML/DL approaches into three principal clusters:
\begin{itemize}
    \item Provision of useful information to assist decision-making, such as human vision attention-guided classification and enhanced model interpretability.
    \item Streamline clinical workflows by aiding in trivial tasks, such as data annotation, detection, and segmentation.
    \item Assessment of expertise, workload, and skills of healthcare professionals.
\end{itemize}
However, these categories are not mutually exclusive, and as we review these papers from different aspects, we see some notable overlaps in the categorization. This interconnection enriches the analysis, allowing a more comprehensive understanding of the themes and findings. The overlap, rather than being a limitation, presents a multifaceted view and amplifies the depth of review.

\subsection{Image Classification} 

Search strategy refers to exploring how medical professionals visually inspect medical images by chasing their gazes around where and how long they look at specific areas of an image during their analysis. Learning from the experts' ability to focus on the most relevant elements in medical images and mimic their search behavior helps to improve the model accuracy for automated image analysis tools. An essential aspect of improving computer-assisted diagnosis concerns operating algorithms to assist clinicians in making more informed and accurate diagnostic decisions. This section highlights the evolving application of eye-gaze data in enhancing the accuracy and efficiency of classifying medical images.

\emph{Akerman et al.(2023)}~\cite{S_006}attempted to detect the glaucoma condition within binary classification models and infuse the ophthalmologist's eye movement strategies to differentiate between expert and novice clinicians.
\emph{Ma et al.(2022)}~\cite{S_008} tackled shortcut learning in deep neural networks, especially in the context of medical image analysis. They proposed a method to rectify harmful shortcut learning in ML/DL models for medical image analysis. They demonstrated an application in chest X-ray classification guided by radiologists' eye-gaze data. Heat maps were generated for individual images based on the radiologists' scan paths and subsequently integrated into a vision transformer (ViT) model in the form of masks, allowing the model to focus on the areas that draw the attention of radiologists. 
\emph{Bhattacharya et al.(2022)}~\cite{S_021} presented RadioTransformer, a framework that integrates radiologists' gaze patterns into a student-teacher transformer model for disease diagnosis in chest radiographs. This approach aims to mimic expert visual search strategies and improve the interpretive capabilities of deep learning models in medical diagnostics. In another research, these authors named GazeRader [4] focuses on integrating radiomics features with visual attention features using a student-teacher framework that is pre-trained and fine-tuned on eye-gaze data obtained from radiologists, and it employs approaches like Radiomics-Visual Attention Loss (RVAL) for effective disease localization in chest radiographs.
Wang et al.~\cite{S_017} aimed to address the limitations of random data augmentation in contrastive learning, using contrastive learning guided by radiologists' visual attention. Their approach emphasizes the importance of guided learning in enhancing the detection and interpretation of critical features in medical images.
\emph{Khosravan et al.}~\cite{PMC_002} in a study aims to address the limitations of existing Computer Aided Detection/Diagnosis (CAD) systems and human error in radiological screenings by integrating eye-tracking data with a CAD system. They hypothesize that combining the strengths of radiologists (good at eliminating false positives) and CAD systems (better at capturing missing tumors than human observers) will improve screening and diagnosis performance.

\subsection{Model Interpretability Enhancement}

The goal of many studies is to explore the potential to improve model interpretability and its performance and validate the hypothesis that human attention can significantly benefit classification models in pathology~\cite{S_011}, mamogram~\cite{S_019}, CXR~\cite{S_005, S_018, S_024, S_025}, ultrasound~\cite{S_022}, and PET scans~\cite{S_014}.
The authors in ~\cite{S_007} focus on understanding how saliency methods for interpreting these different models for chest X-ray classification compare to the eye gaze data of radiologists. They explore the potential to improve model interpretability and its performance.
In research written by Teng et al.,~\cite{S_023}, the core objective of the paper is to enhance disease classification accuracy by understanding and incorporating the visual attention mechanisms of radiologists. The approach is particularly adapted to categorizing diseases in chest radiographs.
Finally, analysis in~\cite{S_024} explores the impact of integrating radiology images, associated report text, and radiologist eye-gaze data on the performance of classification systems and the development of explainable models in medical imaging. This comprehensive approach aims to enhance the accuracy and interpretability of diagnostic tools in radiology.

\subsection{Data Annotation}
Eye-tracking technology enhances the data annotation process by highlighting the search patterns of experts during the meticulous labeling of medical data (including images, text, and genetic information). This technology demands not only high accuracy and domain-specific knowledge but also high accuracy, domain knowledge, and ethical consideration.
Few studies investigated the feasibility of using eye gaze tracking data for labelling medical data. The study by Mariam et al. (2022) focused on utilizing smart gaze-based annotation to label histopathology images. The hand-labeled annotation methods in this paper derived from eye-tracking data and utilized Kernel Density Estimation (KDE) to refine the raw gaze data into more precise annotations by filtering out noise and enhancing the accuracy of the detected ROIs, which offers a promising alternative to traditional manual methods.
The study by Teng et al. (2022) focused on facilitating the analysis of how sonographers navigate and prioritize different anatomical features during scans, offering insights into their skill and technique. Their annotations method specifically involved identifying landmarks that experts focused on while scanning for other anatomical planes, such as the brain and heart, highlighting the complexity and variety in gaze patterns, and the process involved normalizing eye-tracking data to the anatomy circumference using affine transformer networks and then visualizing this data through time curves to characterize and differentiate scanning patterns based on the identified landmarks.

\subsection{Pathology Detection and Segmentation} 
Using the eye-gaze tracking data is a unique supervision format to train the DL/ML-based approaches in object detection and provide the object's precise location within an image. A few studies focused on object detection and localization with eye gaze data. The first study, \emph{GazeRader}~\cite{S_013}, combined radiologists' eye-gaze patterns and radiomics features from chest radiographs for disease localization. The model generates a joint performance evaluation on considerable datasets and demonstrates superior performance in disease localization and classification tasks across multiple large-scale chest radiograph datasets. The second study ~\cite{S_016} concerns improving diagnostic accuracy by integrating eye-tracking data, combining fixation maps with CXR images, and performing abnormality detection for five classes. The third study ~\cite{S_020} explores the viability and timing comparisons of eye gaze labeling compared to conventional manual labeling for training object detectors. Another study ~\cite{PMC_005} provides a proof of concept for an algorithm that combines eye tracking and speech recognition to automatically extract lesion location labels for deep learning, achieving a high accuracy level of 92\% in this process. 
An additional study ~\cite{PMC_006}  addressed the issue of annotating the bounding boxes in CXR datasets, which are costly to collect. This study utilized eye-tracking data recorded from radiologists during their clinical workflow, and by linking this data with the dictation of keywords in radiology reports, tried to supervise the CNNs for specific abnormality localization. One study focuses on improving contrastive learning in medical imaging, particularly X-rays, by incorporating radiologists' visual attention data ~\cite{S_017}. It aims to demonstrate that a gaze map predictor trained on radiologists' gaze data can outperform commonly adopted methods in accurately predicting areas of interest in medical images. The Last study ~\cite{PMC_005} provides proof of concept for an algorithm that combines eye tracking and speech recognition to automatically extract lesion location labels for deep learning, achieving a high accuracy level of 92 percent in this process.

\subsection{Assess Expertise, Workload and Skills} 
This application involves understanding the human factors in medical image interpretation and diagnosis, such as the impact of fatigue and skill level on diagnostic accuracy. Although much research has been done to analyze the workload and skills of radiologists, this point is sparsely considered with the ML/DL algorithms, or the main concentration is on other applications like classification. \emph{Martinez et al.} in~\cite{S_002} discussed the unreliability of qualitative descriptions of radiologists' search patterns, which can affect the quality of improvement interventions and negatively impact patient care. The study aims to overcome this challenge by employing a strategy to understand and analyze these search patterns more effectively. \emph{Pershin et al.} in two different research ~\cite{PMC_001, S_001} conducted experiments aimed at clarifying the influence of workload and fatigue on the precision of radiological interpretations, reducing errors during the interpretations, as well as the alterations in the search patterns shown by radiologists. Different factors can impact the accuracy and efficiency of radiological assessments, and addressing these issues is essential for improving the overall quality and reliability of radiological evaluations.  
Traditional eye-tracking methods primarily focus on the physical dimensions of gaze, such as fixation location and movement patterns, and often overlook the underlying semantic context of the viewer's focus. This focus on quantifiable metrics rather than the interpretative aspects of gaze analysis is highlighted in initial works such as \emph{Noton and Stark's 1971} exploration of scan paths in reading and picture perception, which laid foundational principles for understanding eye movements in terms of spatial patterns without delving deeply into the semantic content those patterns might represent ~\cite{hist_last}.
This approach can lead to reliance on subjective manual annotations and a limited understanding of the viewer's cognitive processing. \emph{Castner et al.}~\cite{S_004} focus on modelling eye movements to distinguish expertise behaviour. This method aims to classify expertise levels by evaluating gaze data from both expert and novice dentists. Research in ~\cite{S_017} highlights the effectiveness of lesion detection in mammograms. The study reveals a strong correlation between where radiologists focus their gaze and the occurrence of diagnostic errors in their evaluations.
Fatigue, high workload, lack of experience or distractions ~\cite{S_009}  and skill ~\cite{S_022, S_023}, and insufficient training lead to misdiagnoses or overlooked conditions. Analyzing eye gaze patterns can assist in identifying areas where radiologists may need further training and recognizing strategies to improve diagnostic accuracy and gain insight into common errors and how to prevent them ~\cite{S_024}.
\section{Visual Attention Learning}
Different visualization approaches assist in interpreting and understanding the images and provide critical insights into how models perceive and process visual information. The principal approaches in the reviewed papers mostly include saliency maps, attention maps, and class activation mapping (CAM) and its variants.

\subsection{Saliency Maps} 
Saliency maps in medical image analysis perform as a bridge between AI models and clinical practice, offering a visual way to understand the decisions made by AI in healthcare and helping through \emph{i) visual presentation} of critical regions of the image in various modalities by color coding or intensity variations to represent how strongly different regions of an image influence the ML/DL algorithm decisions. \emph{ii) Interpretability and explainability} by providing a visual explanation of the approach and enabling medical experts to understand the reasoning behind the diagnosis or prediction presented by the model. \emph{iii) Identifying Biases or errors} in the analysis of the method, and if a model consistently ignores relevant areas or focuses on irrelevant ones, this could indicate a problem with its training or design. \emph{vi) Aid in Diagnosis} by highlighting areas of interest. Saliency maps specify the attention to anomalies or areas of concern that might require further investigation.

\subsection{Class Activation Map (CAM)} 
CAM and its variants are advanced techniques used in Convolutional Neural Network models to visualize the focused regions of the image, and this generated heatmap overlay highlights the features that contributed most to the model's decision, specifically in tasks such as image classification and object detection. All methods in this family are practical in managing the model behavior and generating the overlaid visual heatmap to improve the interpretability of the CNNs' models. In image processing, attention mechanisms consider weights for different areas of the image to specify the most relevant region to improve the model performance and verify if the model focuses on the proper part.

\subsection{Attention Maps} 
The human visual attention system inspires attention maps in neural network models and is vital in visualizing and understanding the model process, prioritizing different regions in an image. Attention maps are classified as soft, hard, and self-attention. While the first one allows the model to focus on multiple parts of the image simultaneously, the second approach selectively focuses on specific regions while ignoring others, and self-attention permits models to weigh the importance of different regions and enrich contextual understanding. Many studies on eye-gaze tracking focus on attention consistency, aiming to limit the gap between the Visual Attention Models (VAM) and CAM. The first is based on the input eye-gaze data, while the model generates the second one.

\section{Integration of Eye-tracking Data to ML/DL Models}

Systematic visual search patterns improve diagnostic decision-making across multiple disciplines, including radiology, pathology, and surgery~\cite{PMC_001}. However, due to the diverse methodologies and approaches employed in interpreting medical images, the availability of the commonly accepted strategies for image interpretation is undefined~\cite{PMC_004, PMC_006}. This section provides a comprehensive overview of the model architecture grounded in eye-gaze tracking data. Table 2 presents a detailed summary, encompassing datasets used, methodologies employed, the types of loss functions integrated into their approaches, and the performance metrics applied to evaluate their work. Furthermore, the rows with no parameters indicate that the article did not specify this information. Following this, we outline the primary objectives of the papers that utilize ML/DL in the analysis of medical images. We categorized the reviewed papers into four distinct classes based on their core methodologies: studies relying on conventional machine learning algorithms(3\%),  those utilizing convolutional neural network (CNN) Algorithms(65\%), those based on transformer Methods(26\%), and those adopting Graph Neural Network based approaches(6\%).

\subsection{Conventional Machine Learning Algorithms} 
Only one paper investigated various conventional machine learning algorithms in their study. \emph{Teng et al.~\cite{S_023}} focused on gradient-boosting decision trees, which improved prediction by combining weak decision trees into a robust predictor. Specifically, they employed Extreme Gradient Boosting (XGBoost) and CatBoost as classifiers. XGBoost is famous for efficiently handling sparse tabular data with scalability, while LightGBM is adept at processing large datasets with numerous instances and features. CatBoost, on the other hand, excels in managing categorical features effectively.

\subsection{Convolutional Neural Network (CNN)}
CNNs represent a significant branch of deep neural networks known for their effectiveness in computer vision tasks. The architecture of the CNN's model helps to learn different features from images, from simple to complex, which is very useful in various applications. In medical imaging, the interpretability of CNNs is particularly beneficial, allowing for more accurate and detailed analysis of medical images. This interpretability, coupled with the network's ability to handle complex image data, makes CNNs a valuable tool in medical imaging for tasks like disease diagnosis and anomaly detection.

Most of the papers in review present their approach based on CNN's architecture. The \textbf{U-Net}~\cite{PMC_001, PMC_003, PMC_004, PMC_005, S_001, S_018} architecture was developed mainly for biomedical image segmentation tasks, and its unique U-shaped architecture, with two main downsampling and upsampling paths, can be modified and extended to work with various image sizes and different kinds of image data.
\textbf{ResNet}~\cite{S_010, S_014, S_017, S_019, S_025, PMC_001, PMC_006}, with its considerable depth, can address the vanishing and exploding gradient problem that often comes with deeper networks and introduced the concept of residual learning. In ResNet, each layer learns features based on the previous layer's output, and each layer learns residual functions concerning the layer inputs instead of learning unreferenced functions. This architecture in eye gaze tracking applications supports predicting the point of gaze based on eye images or related features.
\textbf{MobileNet}~\cite{S_016}, known for its lightweight architecture optimized for performance on mobile and embedded devices, and \textbf{AlexNet}~\cite{S_002}, a pioneering deep CNN that advanced image recognition in eye tracking, \textbf{YOLO}~\cite{S_020} renowned for its real-time object detection capabilities through a single neural network pass. \textbf{MDFN} or Multi-level Dependency Fusion Network~\cite{S_015}, which is a specialized architecture of the CNNs family designed to incorporate multi-level dependencies and handle data fusion or feature extraction at different levels, \textbf{VGG} ~\cite{S_004}, \textbf{RNN- LSTM}~\cite{S_024}, and a combination of the ResNet with EfficientNet and UNet~\cite{S_001, S_005, S_014} are the most CNN approaches that used in the application of the eye tracking for medical image analysis.

\subsection{Transformer-Based Methods} 
Over the years, attention mechanisms have been utilized in computer vision alongside convolutional networks or as replacements for certain parts of these networks and maintaining their fundamental architecture. Vision Transformer(ViT)~\cite{Other_ViT_001} in 2020 indicates that applying pure transformers directly to sequences of image patches is highly effective for image classification tasks. More precisely, ViTs use self-attention mechanisms similar to those in transformers used for natural language processing, which allows ViTs to focus on different parts of the image and understand the relationships between these parts, leading to better performance in tasks like image classification, object detection, and more. Hence, In medical image analysis, the potential of ViT models on large datasets induces more transformer-based research. In this section, six studies using transformer-based architecture are reviewed.

Eye-Gaze-Guided Vision Transformer(EG-ViT)~\cite{S_008} utilizes the region of interest extracted from the radiologist's eye-gaze data. The process begins by generating a visual attention map based on the eye-gaze data. The original image and the corresponding heatmap are then randomly cropped to a smaller size, dividing the cropped image into several patches for patch embedding. Subsequently, an eye-gaze mask is applied to filter out patches unrelated to the pathology and the radiologists' interests. Finally, these masked image patches are fed as input to the transformer for processing.

Research known as RadioTransformer~\cite{S_021} is an end-to-end framework containing two parallel networks (student and teacher). Each network comprises two global and focal components designed to analyze chest radiographs for disease classification. The global components capture the overall image context, facilitating broad understanding, while the focal components concentrate on specific regions, allowing for detailed analysis. This dual-component structure enables the RadioTransformer to focus on relevant areas, similar to how a radiologist examines images, enhancing its diagnostic precision. This approach is mainly beneficial in medical imaging, where detailed observations are crucial for accurate diagnosis. The following study, Mammo-Net~\cite{S_019}, is another breast cancer diagnostic model research incorporating two key components: a multi-view classification network and an attention consistency module. The multi-view classification network processes mammograms from two views (CC and MLO). The attention consistency module utilizes gaze data from radiologists, providing positional supervision to guide the network's focus on relevant areas in the images. This approach aims to enhance the interpretability and performance of mammogram classification by integrating radiologists' gaze data and interactive information across multiple views. Wang et al.~\cite{S_009} introduce the Cross-Attention Transformer Module (CATM). as a component of their Eye-Guided Dual-Path Network. The model in the Dual-Path Encoder component integrates gaze information with image data, and (CATM) embeds human cognitive information into the network, enabling communication between the network and human semantic perceptions. Multi-Feature Skip Connection (MSC) combines spatial information during down-sampling to preserve segmentation details and provide additional information on organ location and edge. Another study ~\cite{S_003} utilizes a transformer's effectiveness for handling visual and textual data, essential for the system's goal of interpreting and imitating radiologists' focus and decision-making process in CXR diagnosis. It involves segmenting input images into patches and utilizing transformer layers for their analysis. A ViT Adapter generates masks and attention heatmaps, guiding the deeper layers of a pre-trained model. Features from the ViT Adapter are then fused with those from BiomedCLIP, another transformer-based model, to merge visual and textual data. BiomedCLIP employs a 12-layer ViT for visual processing and a 12-layer BERT model for text processing. The last study~\cite{S_022} introduces a specialized Affine Transformer Network (ATN) for localizing anatomy in ultrasound frames, which is essential for standardizing eye-tracking data. It uses time curves to visualize and analyze sonographer gaze patterns to discern skill levels. This approach helps identify distinct visual patterns and landmarks that sonographers focus on during scans. By comparing gaze patterns of more and less experienced sonographers, the study aims to characterize their skills. Ultimately, this research offers insights for enhancing training and proficiency in fetal ultrasound scanning.

\subsection{Graph Neural Network (GNN)} 
The proposed model in GazeGNN ~\cite{S_012}is to develop an efficient, real-time algorithm leveraging from graph neural network for disease classification in chest X-rays that leverages raw eye-gaze data without conversion into VAMs. In the graph, each node represents the patch, gaze, and position embedding feature, and the entity is the edges between nodes, which may indicate some form of relationship or interaction between the entities.

\emph{Akerman et al.~\cite{S_006}}  analyzed eye movement patterns of ophthalmologists with various expertise levels, using eye-gaze metrics to develop binary classification models for disease detection and expertise differentiation. They employed supervised and unsupervised learning methods, focusing on the number of fixations and grouping eye fixation data by OCT report regions. The Sequence Graph Transform (SGT) Embedding was used for feature embedding in both learning tasks to interpret fixation point relationships through similarity computations. Finally, they utilized a Feed-Forward Neural Network Model for the supervised learning task, enhancing model accuracy and effectiveness.

\newpage{
\centering
\scriptsize
\renewcommand{\arraystretch}{1} 

\begin{longtable}{
  p{1.5cm}
   >{\centering\arraybackslash}p{3cm }
   >{\centering\arraybackslash}p{3.5 cm}
   p{3.2cm }
   p{4cm}
}

\caption{\small The performance table presents a detailed summary, encompassing datasets used, methodologies employed, the types of loss functions integrated into their approaches, and the performance metrics applied to evaluate their work.}\\

\toprule

\textbf{Reference} &
  \textbf{Dataset} &
  \textbf{Main Method} &
  \textbf{Loss Function} &
  \textbf{Performance Metrics}\\

\midrule
\endfirsthead

\multicolumn{5}{c}%
{{\bfseries \tablename\ \thetable{} -- continued from previous page}} \\
\toprule
\textbf{Reference/Name} &
  \textbf{Dataset} &
  \textbf{Main Method} &
  \textbf{Loss Function} &
  \textbf{Performance Metrics}\\
\midrule
\endhead

\midrule
 \\
\endfoot
\bottomrule
\endlastfoot

\multicolumn{5}{c}{\cellcolor[HTML]{91D9CA}\parbox[c][3em][c]{\linewidth}{\centering\textbf{ \scriptsize Conventional Machine Learning }}} \\
\\

~\cite{S_023} &
  Private &
  \begin{tabular}[c]{@{}c@{}}LightGBM, XGBoost,\\ CatBoost\end{tabular} &
  \_\_ &
 F-Scores 
 \\
  \\ 

\multicolumn{5}{c}{\cellcolor[HTML]{91D9CA}\parbox[c][3em][c]{\linewidth}{\centering\textbf{ \scriptsize Convolutional neural network }}} \\
\\ 
~\cite{S_001} &
VinDr-CXR &
U-Net &
Binary cross-entropy, Dice coefficient losses &
F-Score, Pearson Correlation Coefficient, Wilcoxon Signed-Rank Test, Digit Symbol, Substitution Test, Circle Coverage Test
\\
\\  
~\cite{S_002} & 
Private & 
AlexNet & 
Cross-entropy & 
AUC, Accuracy 
\\
\\  
~\cite{S_004} &
Private &
VGG-16 &
Dissimilarity score between image patches&
Accuracy, Cohen's Kappa scores 
\\
\\
~\cite{S_005} &
CUB-GHA &
ResNet-50, EfficientNet-b5 &
Cross-entropy &
Information Gain, Grad-CAM, Correlation Coefficient, Kullback-Leibler Divergence 
\\
\\
~\cite{S_007} &
MIMIC, EyeGaze CXR &
Gradient,Perturbation, Activation-Based Methods &
Binary Cross Entropy, Mean Squared Error &
Structural Similarity Index score, AUROC 
\\
\\
~\cite{S_010} &
Knee X-ray ,OAI &
ResNet &
Cross-entropy, Attention Consistency, Mean Squared Error&
Accuracy 
\\
\\  
~\cite{S_011} &
Multi-modal CXR &
ResNet50 &
Selective Mean Square Error, Cross Entropy &
AUC, Accuracy 
\\
\\
~\cite{S_013} &
RSNA , SIIM-FISABIO, COVID-19 Detection, NIH CXR, VinBigData, CXR Abnormalities, Detection&
 YOLO &
Generalised IoU, Mean Squared Error&
MSE, AUC 
\\
\\
~\cite{S_014} &
  ADN &
  ResNet18 &
Categorical Cross-Entropy, Dice Coefficient &
  F-Scores,  Accuracy, Sensitivity, Specificity 
\\
\\
~\cite{S_015} &
  MIMIC-Eye &
  MDF-Net &
  \_\_ &
  Average Precision, Average Recall 
\\
\\
~\cite{S_016} &
  REFLACX &
  Mobilenet &
  \begin{tabular}[c]{@{}c@{}}Bounding Box Loss, \\ Mask Loss\end{tabular} &
  \_\_
\\
\\
~\cite{S_017} &
  Private &
  ResNet50 &
  InfoNCE &
  MAE, Accuracy 
\\
\\
~\cite{S_018} &
  PhysioNet &
  multi-task UNet &
  Cross-Entropy, KLD &
  Histogram similarity, Pearson’s correlation coefficient, AUC, ACC
\\
\\   
~\cite{S_020} &
Oral cell dataset, MSRC &
Faster R-CNN, YOLOv3, YOLOv5&
  \_\_ &
  Precision, Recall, Mean Average Precision, Miss-Rate, Log-Average Miss Rate
\\
\\  
~\cite{S_024} &
  Eye Gaze &
  CNN-RNN, LSTM &
  \_\_ &
  AUC 
\\
\\  
~\cite{PMC_001} &
  JSRT &
  U-Net &
  \_\_ &
  Slope Coefficients, and Average Lung Coverage 
\\
\\
~\cite{PMC_002} &
  Private &
  3D deep multi-task CNN &
Classification Loss, Segmentation Loss&
  Average Dice Similarity Coefficient, Accuracy, Sensitivity at Various FP/Scan Rates, Overall Score
\\
\\ 
~\cite{PMC_003} &
  Private &
  U-Net &
  Negative Dice Coefficient &
  AUC, Dice Similarity Coefficient 
\\
\\
~\cite{PMC_004} &
  DICOM &
  U-Net &
  Binary Cross Entropy with Logits Loss &
  AUC, Confidence Interval 
\\
\\ 
~\cite{PMC_005} &
  BraTS &
  U-Net &
  MAE &
  Accuracy
\\
\\
~\cite{PMC_006} &
  REFLACX &
  Resnet50 &
  Composite Hybrid Loss &
  AUC, Intersection over Union
\\
\\
~\cite{S_025} &
Private &
Resnet &
Sparse categorical cross-entropy &
Accuracy, AUC, Correlation Coefficient, Similarity, Kullback-Leibler divergence
\\
\\
\multicolumn{5}{c}{\cellcolor[HTML]{91D9CA}\parbox[c][3em][c]{\linewidth}{\centering\textbf{ \scriptsize Transformer-Based }}} \\

\\

~\cite{S_008} &
Inbreast, SIIM-ACR&
EG-ViT &
Interpretability-guided Inductive bias Loss&
F1-Score,AUG, Accuracy 
\\
\\ 
~\cite{S_021} &
8 different dataset:  2 pneumonia, 4 Covid-19, 2 for thoraci &
Custom-designed Transformer &
Visual Attention Loss, Categorical Crossentropy Loss&
Recall, F1, ACC, AUC, Precision 
\\
\\  
~\cite{S_001} &
VinDr-CXR &
ResNet blocks + transformers &
Binary cross-entropy, Dice coefficient loss&
AUC , F-score, Pearson correlation coefficient 
\\
\\
~\cite{S_009} &
Synapse &
CATM &
CrossEntropyLoss,  Dice Loss &
Dice Similarity Coefficient, Hausdorff Distance 
\\
\\
~\cite{S_003} &
Private &
CAD: ViT Adapter &
Cross-entropy Loss &
Structural Similarity, Peak Signal-to-Noise Ratio, IoU on Foreground and Background, Frequency Weighted IoU 
\\
\\
~\cite{S_022} &
Private &
Affine Transformer Network &
MSE between the transformed image and the ground truth&
Accuracy, Time within the Area of Interest 
\\
\\
\multicolumn{5}{c}{\cellcolor[HTML]{91D9CA}\parbox[c][3em][c]{\linewidth}{\centering\textbf{ \scriptsize Graph Neural Network}}} \\
   \\
~\cite{S_012} &
EyeGaze &
Graph Neural Net &
Cross-entropy Loss &
AUC, Accuracy, Precision, Recall, F1-score 
\\
\\
~\cite{S_006} &
Private &
SGT, Feed-Forward NN &
Binary cross-entropy &
AUC, Accuracy 
\\
\\
\bottomrule
\end{longtable}
\normalsize 
}
\newpage

\section{Conclusion}

The extensive growth in the application of ML/DL across numerous healthcare tasks has prompted research into the potential of integrating AI with radiologists' gaze patterns. This collaboration aims to enhance tasks such as lesion detection and feature extraction for more accurate diagnoses and to optimize treatment plans.
This systematic review delivers a comprehensive study of current research on the intersection of analysis with ML/DL algorithms in medical imaging, including the data, applications, visual learning, and methodologies employed in this field.

There are challenges and restrictions in gathering and managing eye movement data, especially in collecting eye-gaze data from radiologists and other medical experts relying on various imaging modalities.
The need for accurate data annotation is undeniable in supervised learning approaches, and precise annotation based on eye gaze data is a complex and time-consuming task. Additionally, understanding the relationship between a radiologist's gaze and real-time processing of this data, besides the need for ground truth and accurate annotation, presents another obstacle that can be affected by the experts' workload and fatigue, ultimately impacting the results. 

Most research focuses on two-dimensional images, yet there is still a gap in using these images to enhance three-dimensional models.
There is a notable difference between ML/DL methods used for analyzing medical images and those used for analyzing eye-tracking data. Despite its importance, the latter has yet to be extensively explored, indicating a gap in the research.

While many studies have investigated the ability of human visual search patterns and traditional methods to detect lesions across various settings, there remains significant potential for exploration into how ViTs can enhance accuracy beyond conventional approaches. Moreover, incorporating ViTs into clinical workflows and comparing their effectiveness with human visual search patterns in the diagnostic process underscores an essential area for further research.
Many studies utilize eye-tracking data to compare human attention patterns to the model's awareness. This architecture, which relies on attention consistency, employs eye-gaze data in training and validation and omits it in the testing phase, which could compromise the classification accuracy and robustness of the model.
Regarding the operating eye-tracking data, many studies focus on the similarity of human attention patterns and the model's attention, a well-known attention consistency architecture. This architecture, which relies on attention consistency, employs the eye-gaze data in training and validation and omits it in the testing phase, which could compromise the classification accuracy and robustness of the model.
Conversely, the two-stream architecture employs separate units to process eye-gaze information and image data separately. While this method ensures a comprehensive analysis by distinctively handling each data type, it tends to be a time-intensive process, posing challenges in terms of efficiency.
Finally, research has focused on the cognitive aspects of analyzing eye-gaze data for decades. However, in the era of ML/DL, there is a need for increased research efforts to integrate eye-gaze data with forms of expert knowledge, such as clinical notes and diagnostic criteria, further enriching the training process through a multimodal approach.

\medskip

\textbf{Supplementary Materials}:

\textbf{Author Contributions:} 
Conceptualization: S.M., S.L., A.B., and A.D.I; 
investigation: S.M., M.T;
writing original draft preparation: S.M.;
writing review and editing: S.M., S.L., R.A.N, A.D.I; 
visualization: S.M., S.L., and A.D.I; 
validation: S.M., M.T.; 
supervision: S.L., A.B., and A.D.I.
project administration: A.D.I.
All authors have reviewed and consented to the publication of this manuscript's final version.
Funding: This research received no external funding.

\medskip

\textbf{Acknowledgments:} We extend our sincere gratitude to all Computational NeuroSurgery (CNS) Lab members at Macquarie University for their invaluable suggestions and support.
\medskip

\textbf{Conflicts of Interest:}
All the authors declare the absence of any conflicts of interest.
\medskip


\bibliography{references}
\bibliographystyle{abbrv}

\end{document}